\documentclass[aps,prc,twocolumn,superscriptaddress,preprintnumbers]{revtex4-1}
\usepackage[normalem]{ulem}  
\usepackage[dvips]{color} 

\renewcommand\sout{\bgroup \color{red} \ULdepth=-.5ex \ULset}

\usepackage[dvips]{graphicx,color}
\usepackage{amsmath}
\usepackage{amsmath,amssymb,amsfonts}
\usepackage{booktabs}

\newcommand{\tr}{\mbox{tr}}

\newcommand{\sbrace}[1]{\left( #1 \right)}
\newcommand{\mbrace}[1]{\left\{ #1 \right\}}
\newcommand{\bbrace}[1]{\left[ #1 \right]}

\newcommand{\Slash}[1]{\ooalign{\hfil/\hfil\crcr$#1$}}
\newcommand{\Nucl}[2]{$^{#1}$#2}

\newcommand{\DLNucl}[2]{$^{#1}_{\Lambda\Lambda}$#2}

\newcommand{\beq}{\begin{eqnarray}}
\newcommand{\eeq}{\end{eqnarray}}

\newcommand{\btau}{\boldsymbol{\tau}}

\newcommand{\rhoB}{\rho_{\scriptscriptstyle B}}

\def\nuc#1#2{\relax\ifmmode{}^{#1}{\hbox{#2}}\else${}^{#1}$#2\fi}
\newcommand{\comments}[1]{}

\newcommand{\MeV}{\mathrm{MeV}}
\newcommand{\SUv}{SU$(3)_{\mbox{v}}$}

\begin{document}

\preprint{
 YITP-14-11
}

\title{
Isovector potential of $\Sigma$ in nuclei and neutron star matter
}


\author{K.~Tsubakihara}
\email[]{tsubaki@oecu.jp}
\affiliation{
Department of Physics, Faculty of Science,
Hokkaido University, Sapporo 060-0810, Japan.}
\affiliation{
Department of Engineering Science, Faculty of Engineering,
Osaka Electro-Communication University, Neyagawa 060-0810, Japan.}
\author{A.~Ohnishi}
\affiliation{
Yukawa Institute for Theoretical Physics,
Kyoto University, Kyoto 606-8502, Japan.}
\author{T.~Harada}
\affiliation{
Department of Engineering Science, Faculty of Engineering,
Osaka Electro-Communication University, Neyagawa 060-0810, Japan.}


\date{\today}

\begin{abstract}
We determine the coupling constants of $\Sigma$ hyperon with mesons
in relativistic mean field (RMF) models
using $\Sigma^-$ atomic shift data
and examine the effects of $\Sigma$ on the neutron star maximum mass.
We find that we need to reduce 
the vector-isovector meson coupling with $\Sigma$ ($g_{\rho\Sigma}$)
from the value constrained by the \SUv\ symmetry
in order to explain the $\Sigma^-$ atomic shifts
for light symmetric and heavy asymmetric nuclei simultaneously.
With the atomic shift fit value of $g_{\rho\Sigma}$,
$\Sigma^-$ can emerge in neutron star matter
overcoming the repulsive isoscalar potential for $\Sigma$ hyperons.
Admixture of $\Sigma^-$ in neutron stars is found
to reduce the neutron star maximum mass slightly.
\end{abstract}


\maketitle

\section{Introduction}

Neutron star matter equation of state (NS-EOS)
including hyperons is one of the most interesting current subjects 
in nuclear physics as well as in astrophysics.
Hyperons are expected to emerge as the substitutes of nucleons
to reduce the Fermi energy in $\beta$-equilibrium dense matter,
and NS-EOS is strongly affected by the properties of
baryon-baryon interactions~\cite{GD95, Schaffner:1995th, IOTSY, GS98, BM05}:
hyperon-nucleon ($YN$), hyperon-hyperon ($YY$)
and nucleon-nucleon ($NN$) interactions.
Since hyperon lifetimes are too short 
to determine $YN$ and $YY$ interactions precisely via scattering experiments,
we have to deduce information on these interactions through
experimental
~\cite{LambdaData, Nagara, SigmaData, SigmaAtom, XiData}
and theoretical~\cite{Dover:1985ba, Millener:1988hp, Bando:1990yi, SigmaReac}
investigations of hypernuclei
which include one or more $\Lambda$, $\Sigma$ and $\Xi$ hyperons.
NS-EOSs including hyperons have been proposed so far by taking experimental
hypernuclear data into account; they generally predict maximum masses
of neutron stars in the range $(1.3-1.7)\ M_\odot$.
Recent discoveries of the two-solar-mass neutron stars~\cite{NSmass2, NSmass3}
have cast doubt on these EOSs.
The observation is based on the Shapiro-delay,
a consequence of the general relativity,
and the signal is clearly seen 
owing to the fortunate inclination angle ($\sin i \sim 1$).
From this observation, it is concluded that
typical NS-EOSs with hyperons or boson condensates are ruled out.
It is a big challenge to construct NS-EOS
which is consistent with hypernuclear physics results
and supports the two-solar-mass neutron star.

In solving the two-solar-mass NS puzzle mentioned above,
there are two key ingredients:
constraining $YN$ and $YY$ interactions
and understanding the origin of repulsive interactions at high density.
Among $YN$ interactions,
$\Lambda N$ interaction is relatively well-known including its spin dependence,
and we here concentrate on the $\Sigma N$ interaction.
Since $\Sigma^-$ is the lightest among the negatively-charged baryons,
its appearance is favored in neutron stars
because of the charge chemical potential and the nuclear symmetry energy.
For example, Glendenning suggested that $\Sigma^-$ would appear
at $(2-3) \rho_0$ in neutron star matter
in a relativistic mean field (RMF) framework~\cite{SW86,Glendenning1982},
where $\Sigma$ potential in nuclear matter was considered to be similar
to that of $\Lambda$,
$U_\Sigma(\rho_0) \sim U_\Lambda(\rho_0) \simeq -30~\mathrm{MeV}$.
Later on, $\Sigma$ potential in symmetric nuclear matter
is suggested to be repulsive from $\Sigma^-$ atomic shift data~\cite{Mares},
and is confirmed to be repulsive in the quasi-free $\Sigma^-$ production
data~\cite{SigmaData,SigmaReac}.
The $\Sigma N$ repulsion is explained
naturally as a consequence of the quark Pauli blocking
in quark models~\cite{Oka:1986fr,fss2}.
In neutron star matter,
Balberg and Gal pointed out that baryon composition
is sensitive to the choice of the $\Sigma N$ interaction~\cite{BalbergGal},
and similar conclusions are obtained
in RMF approaches~\cite{Schaffner:1995th,IOTSY,Sahu:2001um}.

Now it is commonly understood that the isoscalar part
of the $\Sigma$ potential
is so repulsive that $\Sigma$ hyperons tend to be 
suppressed in NS matter,
while we still have ambiguities in the isovector part of the $\Sigma$ potential.
Typical isovector coupling of $\Sigma$ in RMF is twice that of the nucleons,
$g_{\rho\Sigma} \simeq 2 g_{\rho N}$, owing to the isospin of $\Sigma$,
$I_\Sigma = 2 I_N = 1$.
Atomic shift data of $\Sigma^-$ atoms, however,
suggest much smaller isovector coupling.
From the Si and Pb $\Sigma^-$ atomic shifts,
Mares, Friedman, Gal and Jennings obtained the coupling ratio
$g_{\rho\Sigma}/g_{\rho N} \simeq 2/3$~\cite{Mares}.

In our previous work, we obtained a further smaller ratio,
$g_{\rho\Sigma}/g_{\rho N}=0.434$,
in an RMF model with a logarithmic chiral potential 
motivated by the strong coupling limit of
lattice QCD~\cite{SCLLQCD}
and $\sigma\zeta$ mixing effects from U$(1)_\mathrm{A}$ anomaly~\cite{SCL3}.
In this RMF model, abbreviated as SCL3,
most of the coupling constants have been
constrained by the flavor SU$(3)$ (\SUv) symmetry 
for the vector couplings and experimental data of
nuclear matter, normal and $\Lambda$ hypernuclei and $\Sigma$ atom data.
One exception is the $\rho$-$\Sigma$ coupling;
in order to reproduce atomic shift data of $\Sigma^-$ atoms,
we need to modify $g_{\rho\Sigma}$ from
the \SUv-constrained value.
Smaller $g_{\rho\Sigma}/g_{\rho N}$ ratio leads to 
a less repulsive potential of $\Sigma^-$ in neutron star matter,
and $\Sigma^-$ is found to appear in neutron stars
even though the isoscalar part of the $\Sigma$ potential is repulsive.
The above conclusion, $\Sigma^-$ would appear in neutron stars
with smaller isovector coupling fitting the atomic shifts,
may be model dependent, and should be confirmed
with other RMF model parameters.

Another important aspect for the two-solar-mass NS puzzle
is the origin of the repulsion at high density.
It is well-known that with the non-relativistic effective interaction
derived from the bare two-body $NN$ interaction (g-matrix),
the saturation point depends on the strength
of the tensor interaction and forms a so-called "Coester line",
which is off the empirical saturation point.
When we include phenomenological three-nucleon repulsion
together with the three-body attraction with $\Delta$ in the intermediate state,
it becomes possible to explain the saturation point
and to support two-solar-mass NSs.
The above three-nucleon interactions are, however,
not enough to support heavy neutron stars,
when hyperons are included~\cite{Baldo:1999rq};
the calculated maximum mass of neutron stars with hyperons~\cite{Baldo:1999rq} 
is less than the precisely measured mass of the Hulse-Taylor pulsar,
$1.44 M_\odot$~\cite{Hulse:1974eb}.
We need to introduce three-baryon repulsion,
which also acts in $YNN$, $YYN$ and $YYY$ channels~\cite{Nishizaki:2002ih}.
In a relativistic framework,
three-body repulsion appears naturally from relativistic kinematics.
The attraction from the scalar field appears as the mass reduction,
and its effects are relatively smaller at high densities
compared with the repulsion from the vector field.
As a result, 
the relativistic Br{\"u}ckner-Hartree-Fock (RBHF) theory
can reproduce the saturation point~\cite{DBHF},
and RMF models generally predict large maximum masses of NSs.
%
%
This relativistic repulsion could be enough to explain $1.44M_\odot$,
but it is not sufficient to describe the newly discovered two-solar-mass NSs 
when hyperons are taken into account.
We need to introduce extra repulsion
at high densities also in relativistic frameworks.
One of the mechanisms to get extra repulsion in hyperonic matter
is to introduce three-baryon interaction~\cite{Nishizaki:2002ih}.
Another way may be to introduce repulsive interaction
having different flavor dependence from that adopted 
in current treatments.
The atomic shift fit of $\Sigma^-$ atoms leads to the modification 
of hyperon-meson couplings and is related to the second way.

In this article, we revisit $\Sigma$ hyperons in RMF models
and discuss the possibility of $\Sigma^-$ admixture in neutron star matter.
We compare the results of several RMF models with non-linear meson
self-energies, where hyperon-meson couplings are determined
by reproducing known hypernuclear data.
Especially, we examine whether $\Sigma^-$ should emerge in NS medium
when we adopt the parameter sets which can explain the observed
$\Sigma^-$ atomic shifts.
Finally, we investigate the maximum mass of NS
with NS-EOS constrained by the hypernuclear 
and exotic atom physics requirements.

\section{Relativistic Mean Field including Hyperons}

\subsection{RMF Lagrangian}

RMF models are successful in describing various properties of normal nuclei
with $\sigma$, $\omega$ and $\rho$ mesons which couple with nucleons.
An RMF Lagrangian for normal nuclei and nuclear matter is given as
\begin{align}
\mathcal{L}_N =
& \sum_{i\in N}\bar\psi_i \left( i \Slash{\partial} - M_i
      \right) \psi_i
	+ \mathcal{L}_{\sigma\omega\rho}
\nonumber\\
   +& \sum_{i\in N}\bar\psi_i \left[
	g_{\sigma i} \sigma
	- \gamma_\mu (g_{\omega i} \omega^\mu
	+ g_{\rho i} \btau\cdot\mathbf{R}^\mu )
      \right] \psi_i
\label{Eq:RMFLagN}
\ ,\\
\mathcal{L}_{\sigma\omega\rho} =
  & \frac12\,\partial_\mu\sigma\partial^\mu\sigma
                    - \frac12\,m_\sigma^2 \sigma^2
   - V_{\sigma}\left(\sigma \right)
\nonumber\\
   -& \frac14\omega_{\mu\nu}\omega^{\mu\nu}
   + \frac{m_\omega^2}{2} \omega_\mu\omega^\mu
   + V_{\omega}\left(\omega \right)
\nonumber\\
  -& \frac14     \mathbf{R}_{\mu\nu}\cdot\mathbf{R}^{\mu\nu}
   + \frac{m_\rho^2}{2}    \mathbf{R}_\mu\cdot\mathbf{R}^\mu
\ ,
\end{align}
where
$V^{\mu} (V=\omega, \mathbf{R})$ shows
the field tensor of the $\omega$ or $\rho$ vector mesons,
and $\btau$ represents the isospin Pauli matrix.
$V_\sigma$ and $V_\omega$ represent 
$\sigma$ and $\omega$ self-energies,
\begin{align}
V_\sigma(\sigma) = \frac13\, c_{\sigma 3} \sigma^3 +
	                           \frac14\, c_{\sigma 4} \sigma^4
\label{Eq:NLterm}
\ ,\\
V_\omega(\omega) = \frac14\,c_{\omega 4}\sbrace{\omega_\nu\omega^\nu}^2
\ .
\end{align}
We adopt here NL1~\cite{NL1}, NL-SH~\cite{NL-SH} and TM1~\cite{TM1}
as typical RMF models for normal nuclei.
We also examine the former SCL model (SCL2)~\cite{SCL2},
where the $\sigma$ self-energy was derived from analytical
calculation in the strong coupling limit of lattice QCD and reads
\begin{gather}
V_\sigma(\sigma) = 
	- \frac{f_\pi^2\left(m_\sigma^2 - m_{\pi}^2\right)}{2}
\left[  \log \left(1 - \frac{\sigma}{f_\pi}\right)
       + \frac{\sigma}{f_\pi}
       + \frac{\sigma^2}{2f_\pi^2}\right]
	\ .
	\label{Eq:log}
\end{gather}
Their parameter sets are summarized in Table \ref{tab:prmNLRMF}.
We note that coupling constants of these mesons and nucleons
are well constrained by fitting binding energies of normal nuclei,
while non-linear meson self-energy terms are not determined precisely.
While these sophisticated RMF models describe normal nuclear properties well,
differences in non-linear terms give rise to large ambiguities
in dense matter EOS.
Thus it would be possible to discriminate these RMF models for normal nuclei
by including hyperons and applying them to NS-EOS.

RMF has been extended to describe also hypernuclei
and hyperonic matter~\cite{GD95, Schaffner:1995th, IOTSY, GS98, BM05, SCL3}.
%
%
A simple extension is to include hyperons in the baryon sum
in Eq.~(\ref{Eq:RMFLagN}), $i\in B$,
where $B$ represents nucleons ($N$) and hyperons ($Y$).
It is more natural to include $\zeta$ and $\phi$,
scalar and vector mesons consisting of $\bar{s}s$, respectively,
which generate additional attractive and repulsive interactions
among hyperons.
A typical RMF Lagrangian including hyperons, $\zeta$ and $\phi$ mesons
is given as,
\begin{align}
\mathcal{L}_B =&
 \sum_{i\in B}\bar\psi_i \left(i \Slash{\partial} - M^*_i + \gamma_\mu V^\mu_i
 \right) \psi_i + \mathcal{L}_{\sigma\omega\rho} + \mathcal{L}_{\zeta\phi}
\ ,
\label{Eq:RMFLag}
\\
\mathcal{L}_{\zeta\phi} =
   & \frac12\,\partial_\mu\zeta\partial^\mu\zeta 
                    - \frac12\,m_\zeta ^2 \zeta ^2
   - \frac14\phi_{\mu\nu}\phi^{\mu\nu}
   + \frac{m_\phi^2}{2} \phi_\mu\phi^\mu
\ , \label{Eq:RMFLagH}
\end{align}
where
$\phi^{\mu\nu}$ is the field field tensor for $\phi$.
The baryon effective masses $M^*_i$ and the vector potentials $V^\mu_i$
are given as,
\begin{align}
M^*_i =& M_i + S_i\ ,\label{Eq:EffMass}\\
S_i =& -\left(g_{\sigma i}\sigma + g_{\zeta i}\zeta\right)\ , 
\label{Eq:Scalar}\\
V^\mu_i =& g_{\omega i}\omega^\mu + g_{\rho i} \btau\cdot\bold{R}^\mu
	        + g_{\phi i}\phi^\mu\ .
\label{Eq:Vector}
\end{align}
Here $\btau$ represents the isospin Pauli matrices for $I=1/2$ baryons
($N$ and $\Xi$),
and the isospin matrices for $I=1$ ($\Sigma$) baryon.
In order to keep the normal nuclear properties
in the original RMF models,
we assume that nucleons do not couple with $\bar{s}s$ mesons 
and we set $g_{\zeta N}=g_{\phi N}=0$.
This treatment also means that we respect the OZI rule~\cite{OZI},
where $\bar{s}s$ does not couple with nucleons,
\textit{i.e.,} hair-pin diagrams are suppressed.

\begin{table}[tbh!]
\centering
\caption{The parameters in RMF models,
NL1~\cite{NL1}, NL-SH~\cite{NL-SH}, 
TM1~\cite{TM1},
and
SCL2~\cite{SCL2}.
The parameters for normal nuclear systems (upper part)
are determined in original references.
The coupling constants between mesons and hyperons (lower part)
are determined in this article based on hypernuclear data.
For comparison, we also show parameters in SCL3 RMF model~\cite{SCL3}
whose meson-hyperon coupling constants have been fixed
in the same procedure as this work.
}
\label{tab:prmNLRMF}
\begin{tabular}{c|ccccc}
\hline
	& NL1		& NL-SH		& TM1		& SCL2		& SCL3	\\
\hline
$M_N$ (MeV)
	& 938		& 939		& 938		& 938		& 938	\\
$m_\sigma$ (MeV)
	& 492.250	& 526.059	& 511.198	& 502.63	& 690	\\
$m_\omega$ (MeV)
	& 795.359	& 783		& 783		& 783		& 783	\\
$m_\rho$ (MeV)
	& 763		& 763		& 770		& 770		& 770	\\
$g_{\sigma N}$
	& 10.1377	& 10.444	& 10.0289	& 10.08		& 10.15	\\
$g_{\omega N}$
	& 13.2846	& 12.945	& 12.6139	& 13.02		& 11.95	\\
$g_{\rho N}$
	& 4.9757	& 4.383		& 4.6322	& 4.40 		& 4.54	\\
$c_{\sigma 3}$ (fm$^{-1}$)
	& $-$12.1734	& $-$6.9099	& $-$7.2325	& -		& -	\\
$c_{\sigma 4}$
	& $-$36.2646	& $-$15.8337	& 0.6183	& -		& - 	\\
$c_{\omega 4}$
	& 0		& 0		& 71.3075	& 200		& 294.9	\\
\hline
$m_\zeta$ (MeV)
	& 980		& 980		& 980		& 980		& 826.3	\\
$g_{\sigma\Lambda}$
	& 6.10 		& 6.405		& 6.04		& 6.215		& 3.40	\\
$g_{\zeta\Lambda}$
	& 6.31 		& 5.85		& 5.93		& 5.80		& 5.17	\\
$g_{\sigma\Sigma}$
	& 4.83 		& 5.13		& 4.86		& 4.72		& 3.16	\\
$g_{\rho\Sigma}$
	& 2.48		& 1.85		& 1.87		& 1.67		& 1.97	\\
\hline
\end{tabular}
\end{table}

\subsection{Hyperon-Meson Coupling Constants}

To examine the neutron star matter properties based on the RMF Lagrangian,
Eq.~\eqref{Eq:RMFLag},
we start from fixing the coupling constants of mesons and hyperons:
$g_{\sigma Y}$, $g_{\zeta Y}$, $g_{\omega Y}$, $g_{\rho Y}$, and $g_{\phi Y}$.
Unfortunately, it is so time-consuming and sometimes meaningless 
to vary each coupling constant independently since only their balance
can affect the calculated numerical properties.
Thus, there are mainly two types of 
prescriptions to constrain meson-hyperon coupling sets.
One of them is based on the picture where we regard the mesons
in RMF models are made of $\bar{q}q$.
Then the hyperon-meson couplings are constrained
by symmetries of quarks.
The other is based on the chiral perturbation theory.
Scalar and vector fields are generated by the Nambu-Goldstone bosons
(pions, kaons, and eta) and low energy coefficients,
and $\sigma$ and $\omega$ mesons in RMF are considered
to be effective mesons, which represent the scalar and vector fields
but are not actual mesons.
This picture generally gives smaller $\sigma$-hyperon and $\omega$-hyperon
couplings compared with the former picture.

Many of RMF models adopt the former picture and
assume some symmetry relations in vector meson-baryon coupling constants.
For example, some of RMF models employ SU(6) symmetric coupling constants,
which corresponds to the naive quark counting.
The flavor SU(3) symmetry (\SUv\ symmetry) is known to be 
a better symmetry in hadrons,
and constrains the vector meson-baryon interaction Lagrangian as,
\begin{align}
\mathcal{L}_{\mathrm{BV}}^\mathrm{SU(3)}
= \sqrt{2}\{& g_s\,\tr\sbrace{M_v} \tr\sbrace{\bar{B}B}
              + g_D\,\tr\sbrace{\bar{B}\mbrace{M_v, B}} \nonumber\\
		    & + g_F\,\tr\sbrace{\bar{B}\bbrace{M_v, B}}\}\nonumber\\ 
= \sqrt{2}\{& g_s\,\tr\sbrace{M_v} \tr\sbrace{\bar{B}B}
              + g_1\,\tr\sbrace{\bar{B}M_vB} \nonumber\\
		    & + g_2\,\tr\sbrace{\bar{B}BM_v}\}\ .
	\label{Eq:SUV3}
\end{align}
Here, $B$ and $M_v$ are flavor SU(3) baryon and vector meson matrices.
Under the \SUv\ symmetry with the assumption $g_{\phi N}=0$, all
vector meson-hyperon coupling constants
are constrained once $g_{\rho N}$ and $g_{\omega N}$ are fixed.
From Eq.~(\ref{Eq:SUV3}), \SUv\ vector coupling constants
for $\Lambda$ and $\Sigma$ hyperons are given as
\begin{align}
  &g_{\omega\Lambda}^\mathrm{SU(3)} = \frac{5}{6}g_{\omega N} - \frac{1}{2}g_{\rho N},\;\;
  g_{\phi\Lambda}^\mathrm{SU(3)} = \frac{\sqrt{2}}{6}\sbrace{g_{\omega N} + 3g_{\rho N}}\ ,
	\label{Eq:SUV3Lambda}
\\
  &g_{\omega\Sigma}^\mathrm{SU(3)} = \frac12\sbrace{g_{\omega N} + g_{\rho N}},\ 
  g_{\phi\Sigma}^\mathrm{SU(3)}   = \frac{\sqrt{2}}{2}\sbrace{g_{\omega N} - g_{\rho N}}\ ,
\nonumber\\
  &g_{\rho\Sigma}^\mathrm{SU(3)} = 2 g_{\rho N}\ .
	\label{Eq:SUV3Sigma}
\end{align}
Naive quark counting also follows the above coupling constant relations.
For example, when we set $g_{\rho{N}}=g_{\omega{N}}/3$,
the above relations lead to the quark counting relation,
$g_{\omega\Lambda} = g_{\omega\Sigma} = 2 g_{\omega{N}}/3$.
The remaining scalar coupling constants, $g_{\sigma Y}$ and $g_{\zeta Y}$,
may be fixed by explaining experimental data of hypernuclear systems.

It should be noted that the above relations are based
on the $\bar{q}q$ picture for RMF mesons in the flavor SU(3) limit.
If the mesons in RMF contain significant components
generated by pions or if the SU(3) breaking effects
are strong, hyperon-meson couplings can deviate from the relation
in Eqs.~\eqref{Eq:SUV3Lambda} and \eqref{Eq:SUV3Sigma}.
While it is generally believed that hyperonic EOSs are ruled out
by the two solar mass neutron stars~\cite{NSmass2,NSmass3},
the naive quark counting relation mentioned above
is respected for $\omega$-hyperon ($\Lambda$ and $\Sigma$) couplings,
$g_{\omega{Y}} \simeq 2g_{\sigma{N}}/3 $,
in the ruled-out hyperonic EOS~\cite{GM1991}.
In the original paper by Glendenning and Moszkowski~\cite{GM1991},
however, the authors considered other possibilities
where heavier neutron star can be supported by hyperonic EOS.
Furthermore, additional hyperon-hyperon repulsion coming
from the $\phi$ meson exchange was not considered.
Thus we need more care to set the hyperon-meson couplings.

In this work,
we adopt flavor \SUv\ symmetric couplings 
shown in Eq.~(\ref{Eq:SUV3}) as a starting point,
and modify some of the coupling constants
which have large effects in explaining the hypernuclear data.
This procedure enables us to construct NS-EOS
which includes hypernuclear information effectively.
In the next section, we try to determine the hyperon-meson couplings
based on the experimental data.

\section{Hypernuclei, Exotic Atoms, \\ and Neutron Stars}
\label{Sec:Results}

In this section,
following the procedure adopted in our previous work~\cite{SCL3},
we introduce $\Lambda$ and $\Sigma$ hyperons
in the NL1, NL-SH, TM1, and SCL2 RMF models,
and fit
the $\Lambda$ separation energies data in single $\Lambda$ hypernuclei,
the $\Lambda\Lambda$ bond energy data in the double $\Lambda$ hypernucleus,
and the $\Sigma^-$ atomic shifts data.
Next we apply the obtained RMF model parameters to calculate
the neutron star matter EOS.
We can find similar works in the literature,
but information of $\Sigma^-$ atoms were not taken into account and
stronger $\Lambda\Lambda$ attraction was assumed in~\cite{Schaffner:1995th},
and only one set of the normal nuclear RMF models was used in ~\cite{Mares}.

\subsection{Lambda Hypernuclei}


For $\Lambda$ hyperon,
we have four RMF parameters to be determined:
$g_{\sigma\Lambda}$,
$g_{\zeta\Lambda}$,
$g_{\omega\Lambda}$, and $g_{\phi\Lambda}$.
Experimental data of
$\Lambda$ separation energies ($S_\Lambda$) in single $\Lambda$ hypernuclei
and the $\Lambda\Lambda$ bond energy ($\Delta B_{\Lambda\Lambda}$)
in the double $\Lambda$ hypernucleus
$^{6}_{\Lambda\Lambda}\mathrm{He}$~\cite{Nagara}
are available as the constraints of these coupling constants.
Unfortunately, the $\Lambda$ potential at high density is not very sensitive to
all of these available data.
The baryon potential at low momentum is given
as the sum of scalar and vector potentials,
$U_B = S_B + V_B^0$, 
and this potential mainly determines the hypernuclear properties
at around normal nuclear density.
Both the scalar and vector potentials are approximately proportional
to $\rhoB$ at low density,
thus we cannot determine the scalar and vector potentials
for $\Lambda$ separately.
One may think that additional information on the difference
($S_B - V_B$) is available from the spin-orbit splitting.
However, it is possible to explain the spin-orbit splittings
by tuning the tensor coupling of the vector meson,
which is not incorporated in the RMF Lagrangian considered in this work.
Since the tensor coupling does not affect the EOS of uniform matter
in the mean field approximation, the spin-orbit splitting is not helpful
to constrain the EOS at high density.

\begin{figure}[tb]
	\begin{center}
		\includegraphics[width=0.48\textwidth]{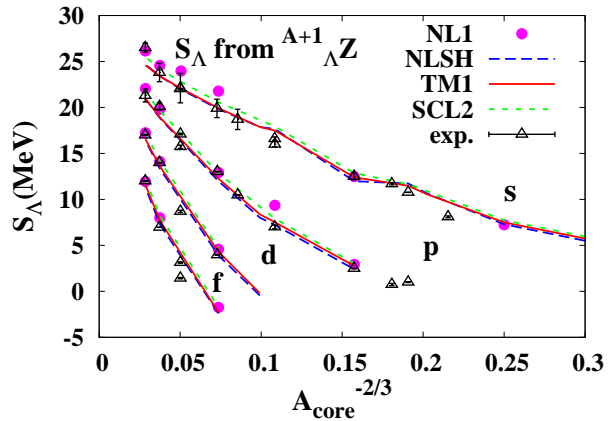}
	\end{center}
	\caption{Experimental and calculated separation energies of $\Lambda$
		from single $\Lambda$ hypernuclei, $S_\Lambda$.
		Magenta point presents the fitting results based on NL1
		RMF model.
		Broken blue lines presents the fitting results based on NL-SH
		RMF model.
		Solid red lines show the fitting results based on TM1
		RMF model.
		Open triangle symbols are experimental $S_\Lambda$.}
	\label{fig:se}
\end{figure}

We adopt here the \SUv\ relations for the vector couplings
in Eq.~\eqref{Eq:SUV3Lambda},
$g_{\omega\Lambda}=g^\mathrm{SU(3)}_{\omega\Lambda}$
and
$g_{\phi\Lambda}=g^\mathrm{SU(3)}_{\phi\Lambda}$
to fix the vector potentials.
The remaining scalar-isoscalar coupling constants,
$g_{\sigma \Lambda}$ and $g_{\zeta \Lambda}$,
are determined by fitting experimental hypernuclear data:
$\Lambda$ separation energies $S_\Lambda$
in single $\Lambda$ hypernuclei
and the $\Lambda\Lambda$ bond energy
$\Delta B_{\Lambda\Lambda}=0.67 \pm 0.17~\MeV$
of the double $\Lambda$ hypernuclei \DLNucl{6}{He}
observed in the NAGARA event~\cite{Nagara,T11}.

The obtained coupling constant sets
($g_{\sigma\Lambda}$ and $g_{\zeta\Lambda}$)
are summarized in TABLE \ref{tab:prmNLRMF}.
In Fig.~\ref{fig:se},
we show the calculated results of $S_{\Lambda}$
in NL1, NL-SH, TM1 and SCL2
by using the obtained parameter sets
as a function of $A_\mathrm{core}^{-2/3}$,
where $A_\mathrm{core}$ is the mass number of the core nucleus.
Since the kinetic energy of $\Lambda$ is approximately proportional
to $1/R^2 \propto A_\mathrm{core}^{-2/3}$,
we can guess the potential depth in nuclear matter as
$U_\Lambda \simeq - 28~\MeV$
from the extrapolation to $A \to \infty$
($A_\mathrm{core}^{-2/3} \to 0$).
We find that experimental $S_{\Lambda}$ values are well explained
in these RMF models.
In addition to the ground state separation energies,
excited single particle energies of $p$, $d$ and $f$ waves are also
well described.
The $\Lambda$ shell gaps reflect the strength of the scalar potential
via the effective mass $M_i^*$ in Eq.~\eqref{Eq:EffMass},
then the scalar potential for $\Lambda$ seems to have an appropriate strength.

\subsection{Sigma-Nuclear Potential and $\Sigma^-$ Atoms}

For $\Sigma$ hyperon,
we have five RMF parameters,
$g_{\sigma\Sigma}$, $g_{\zeta\Sigma}$, $g_{\omega\Sigma}$, $g_{\phi\Sigma}$
and $g_{\rho\Sigma}$.
Because of isospin of $\Sigma$ hyperon,
we have one more parameter 
for the isovector-vector coupling ($g_{\rho\Sigma}$)
compared with $\Lambda$.
%
Since
we have no other knowledge of bound $\Sigma$ hypernuclei other than
${}^4_\Sigma\mathrm{He}$~\cite{He4Sigma},
we have to rely on quasi $\Sigma$ production reactions~\cite{SigmaData}
and $\Sigma^-$ atomic shifts data~\cite{SigmaAtom}.

In $\Sigma^-$ atoms, a $\Sigma^-$ moves around a nucleus
in the Coulomb orbit.
When the $\Sigma^-$ goes down to small $n$ orbit
through subsequent atomic cascade processes
($X$-ray emission or Auger process),
the $\Sigma^-$ is absorbed in the nucleus via the conversion
$\Sigma^- p \to \Lambda n$ inside the nuclei.
The $X$-ray just before the absorption thus contains information 
of $\Sigma^-$-nucleus potential.
$\Sigma^-$ atomic shifts have been measured 
for isospin-symmetric (O, Mg, Al, Si and S)
and heavier isospin-asymmetric (W and Pb) nuclei\cite{SigmaAtom}.
Once we fix the coupling constants of $\Sigma$ with isosinglet mesons,
$\Sigma^-$ atomic shift data in heavy nuclei are useful
to determine isovector coupling, $g_{\rho\Sigma}$.

We fix the $\Sigma$-meson coupling constants in the following way.
First, we obtain core nuclear wave functions in RMF.
Second, $\Sigma^-$-nucleus optical potential is given
as the Schr\"{o}dinger-equivalent potential,
\begin{equation}
	\mbox{Re}V_{\mbox{opt}}^{\Sigma^-} = S_{\Sigma^-}(r) +
	\frac{EV^0(r)}{M_{\Sigma^-}} + 
	\frac{S_{\Sigma^-}^2(r) - (V^0_{\Sigma^-})^2(r)}{2M_{\Sigma^-}}\ ,
	\label{Eq:Sch}
\end{equation}
where $S_{\Sigma^-}$ and $V_{\Sigma^-}^0$ are the scalar
and vector potentials of $\Sigma^-$ hyperon shown
in Eqs.~\eqref{Eq:Scalar} and \eqref{Eq:Vector},
respectively.
Here, the meson fields ($\sigma, \omega, \rho$) used in
$S_{\Sigma^-}$ and $V_{\Sigma^-}^0$ are those of the core nuclei.
Next, we fit the $\Sigma^-$ atomic shifts for light symmetric nuclei
by choosing the isoscalar part of coupling constants properly.
%
%
Since we do not have $\zeta$ and $\phi$ fields in normal nuclei,
atomic shifts have no dependence on $g_{\zeta\Sigma}$ and $g_{\phi\Sigma}$.
We adopt \SUv\ value for $g_{\omega\Sigma}$ and $g_{\phi\Sigma}$,
and we invoke naive quark counting for $g_{\zeta\Sigma}$
and assume $g_{\zeta\Sigma} = g_{\sigma\Sigma}/\sqrt{2}$.
By tuning $g_{\sigma\Sigma}$,
we can well describe $\Sigma^-$ atomic shifts for light symmetric nuclei.

Finally, we determine $g_{\rho\Sigma}$ by fitting the 
atomic shifts of heavier $\Sigma^-$ atoms.
In Ref.~\cite{SCL3}, we found that 
it is difficult to explain the $\Sigma^-$ atomic shifts with 
heavier core-nuclei which is isospin-asymmetric,
if we keep \SUv\ symmetry relation shown in
Eq.~(\ref{Eq:SUV3}).
Thus, we need to modify $g_{\rho \Sigma}$ to reproduce $\Sigma^-$
atomic shift data in the same way as Ref.~\cite{SCL3}.
One of the reasons to modify $g_{\rho\Sigma}$ from the \SUv\ value
may be that the isovector part of $\Sigma$ interaction should be affected
by the quark Pauli principle, which cannot be expressed in the meson
exchange potential with \SUv\ relation.

\begin{figure}[tb]
	\begin{center}
		\includegraphics[width=0.48\textwidth]{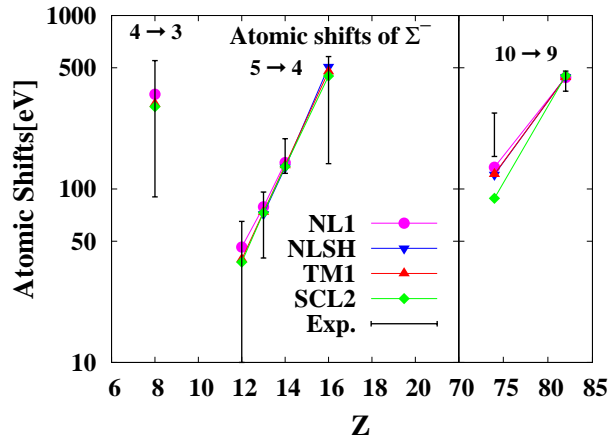}
	\end{center}
	\caption{Experimental and calculated $\Sigma^-$ atomic shifts
		as a function of the atomic number of core nuclei.
		Solid green lines and filled diamonds presents
		the fitting results based on NL1 RMF model.
		Solid blue lines and filled inverse triangles presents
		the fitting results based on NL-SH RMF model.
		Solid red lines and filled triangles show the fitting results
		based on TM1 RMF model.
		Magenta lines are experimental $\Sigma^-$ atomic shifts
		on each core nuclei.}
	\label{fig:AS}
\end{figure}
\begin{figure}[tb]
 	\begin{center}
 		 \includegraphics[width=0.50\textwidth]{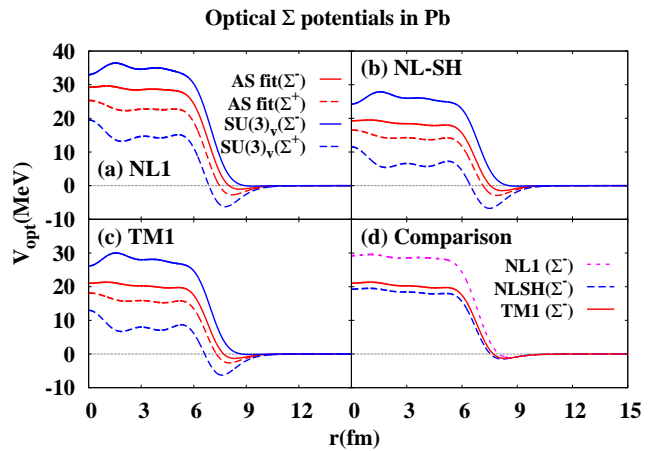}
 	\caption{
		Schr\"{o}dinger-equivalent potentials of
 		$\Sigma^-$ and $\Sigma^+$ calculated with (a) NL1, (b)NL-SH, and
		(c) TM1 RMF models, respectively.
 		The results of parameter sets determined by reproducing 
 		$\Sigma^-$ atomic shift are shown with red lines.
 		The results of parameter sets suggested from SU$(3)_{\mbox{v}}$
		symmetry are presented by blue lines.
 		The real parts of the optical potentials
		of $\Sigma^-$($\Sigma^+$) are
		shown by solid(dashed) lines.
		}
 	\label{fig:potSE}
 	\end{center}
\end{figure}

In Fig.~\ref{fig:AS}, experimental and calculated atomic shifts on several core
nuclei are shown as a function of the atomic number of core nuclei.
Basic trend of $\Sigma^-$ atomic shift is reproduced sufficiently well.
%
Determined $\Sigma^-$-core nuclei optical potentials are shown in
Fig.~\ref{fig:potSE} for NL1, NL-SH, and TM1 models.
Fixed parameters are summarized in Table \ref{tab:prmNLRMF}.
We refer to these fitted $g_{\rho\Sigma}$s as atomic shift (AS) fit values.
Compared to \SUv\ values, $g_{\rho\Sigma} = g_{\omega\Sigma}$,
AS fit values of $g_{\rho\Sigma}$ are strongly reduced.
From these results, it seems common to all employed RMF models that
$\Sigma^-$ feels repulsive potential in nuclear medium
and its height is in the range of $20\sim30$MeV, and that a few MeV
attractive pockets around nuclear surface are essential
to explain experimental $\Sigma^-$ AS especially on \Nucl{208}{Pb}.

In addition to $\Sigma^-$ optical potentials,
we also present $\Sigma^+$ potentials
in both $\Sigma^-$ AS fit and \SUv\ cases.
By comparing $\Sigma^-$ and $\Sigma^+$ potentials,
we can roughly estimate the symmetry energies of $\Sigma$ and their difference
between $\Sigma^-$ AS fit and \SUv\ cases.
The symmetry energies of $\Sigma$ are reduced 
from around 15 MeV with \SUv\ values to around 3 MeV with \SUv\ values
in all employed RMF models.

\subsection{Neutron Star Matter EOS}

Based on the meson-hyperon coupling constants determined
in the previous subsection,
NS-EOS, for example
energy density ($\epsilon=E/V$) and pressure ($P$)
as functions of density,
are obtained from the energy--momentum tensor calculated
by using RMF Lagrangian, Eq.~(\ref{Eq:RMFLag}).
This procedure enables us to deduce reliable NS-EOS which explains
known bulk properties of nuclear and hypernuclear systems.
Here, $\epsilon$ and $P$ are written as 
\begin{align}
\epsilon = T^{00} &= \frac12 m_\sigma^2\sigma^2 + V_\sigma(\sigma)
	           + \frac12 m_\zeta^2 \zeta ^2 \nonumber\\
	          &+ \frac12 m_\omega^2        \omega^2
	           + \frac{3c_{\omega 4}}{4} \omega^4 
                   + \frac12 m_\rho  ^2        \rho  ^2
                   + \frac12 m_\phi  ^2        \phi  ^2 \nonumber\\
		  &+ \sum_{i=B,l}\frac{\nu_i}{(2\pi)^3}\int _0^{k_F^i} d^3k\sqrt{k^2 + (M_i^\ast)^2} \\
P = \frac{1}{3}\sum_i T^{ii}
                  &= - \frac12 m_\sigma^2\sigma^2 - V_\sigma(\sigma)
	           - \frac12 m_\zeta^2 \zeta ^2 \nonumber\\
	          &+ \frac12 m_\omega^2        \omega^2
	           + \frac{c_{\omega 4}}{4} \omega^4 
                   + \frac12 m_\rho  ^2        \rho  ^2
                   + \frac12 m_\phi  ^2        \phi  ^2 \nonumber\\
		  &+ \sum_{i=B,l}\frac{\nu_i}{(2\pi)^3}\int _0^{k_F^i} d^3k\frac{k^2}{3\sqrt{k^2 + (M_i^\ast)^2}}
	\label{Eq.EDP}
\end{align}
In NS matter, the density of each baryons $\rho_{Bi}$ should be determined
under the charge neutrality and the $\beta$--equilibrium conditions.
Thus, the total baryon densities $\rho_B$, the lepton densities $\rho_l$,
the charge density $\rho_c$, and the chemical potential $\mu_{Bi}$ obey
the following equations,
\begin{gather}
	\rho_B    = \sum_{i=B} \rho_{Bi}\\
	\rho_{Bi} = \frac{1}{3\pi^2}\left\{
	            \left( \mu_{Bi} - V_{i}\right)^2 - (M_{i}^\ast)^2 
		    \right\}^{3/2}\\
	\mu_{Bi}  = \mu_B + q_i\mu_c\\
	\rho_c    = \sum_{i=B} q_i\rho_{Bi} + \sum_{j=l} \rho_{lj} = 0
	\label{test}
\end{gather}
This condition means that all reactions are allowed as long as
charge and baryon numbers are conserved. 
For example, $\Lambda$ hyperon can emerge as a substitute of $n$
in the high $\rhoB$ region.
%
%
In this subsection, we examine NS-EOS derived from the RMF models whose
coupling constants between mesons and hyperons have been determined 
in the previous subsection.

\begin{figure}[tb]
	\begin{center}
		\includegraphics[width=0.48\textwidth]{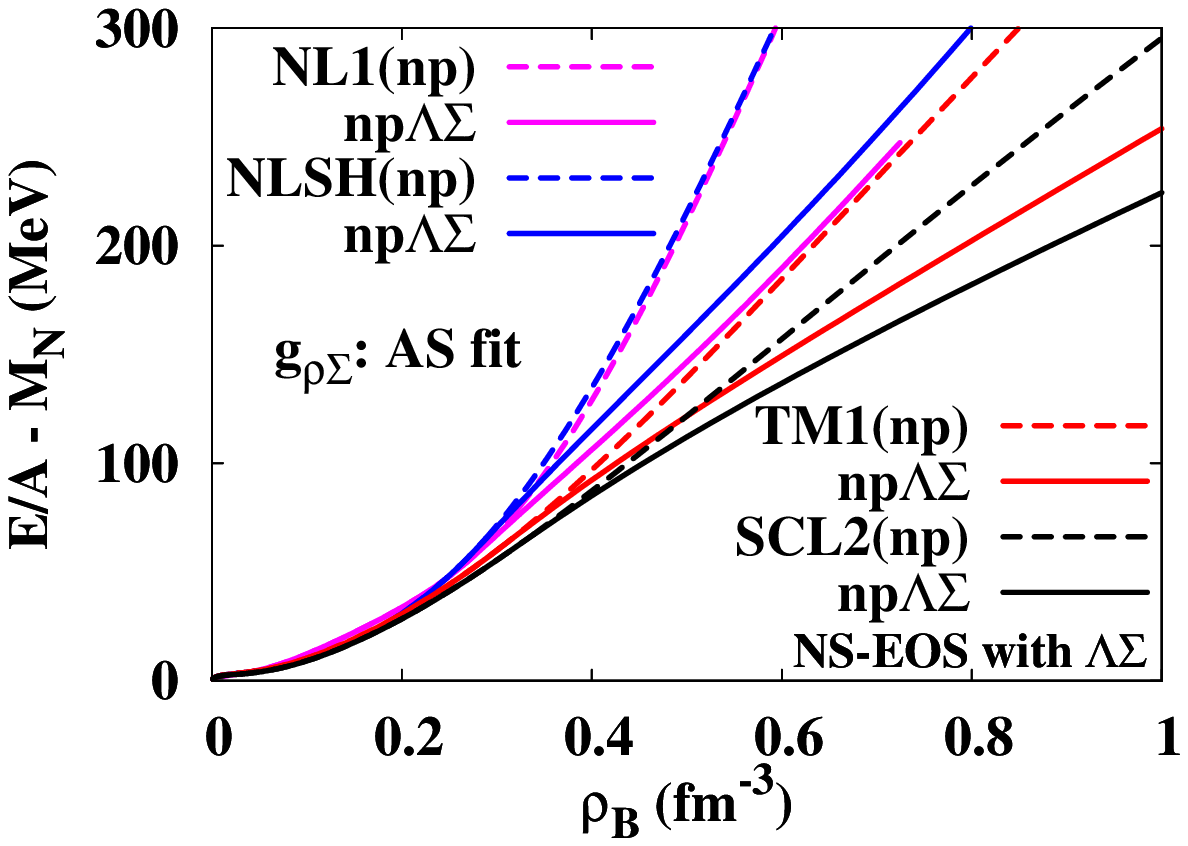}
		\includegraphics[width=0.48\textwidth]{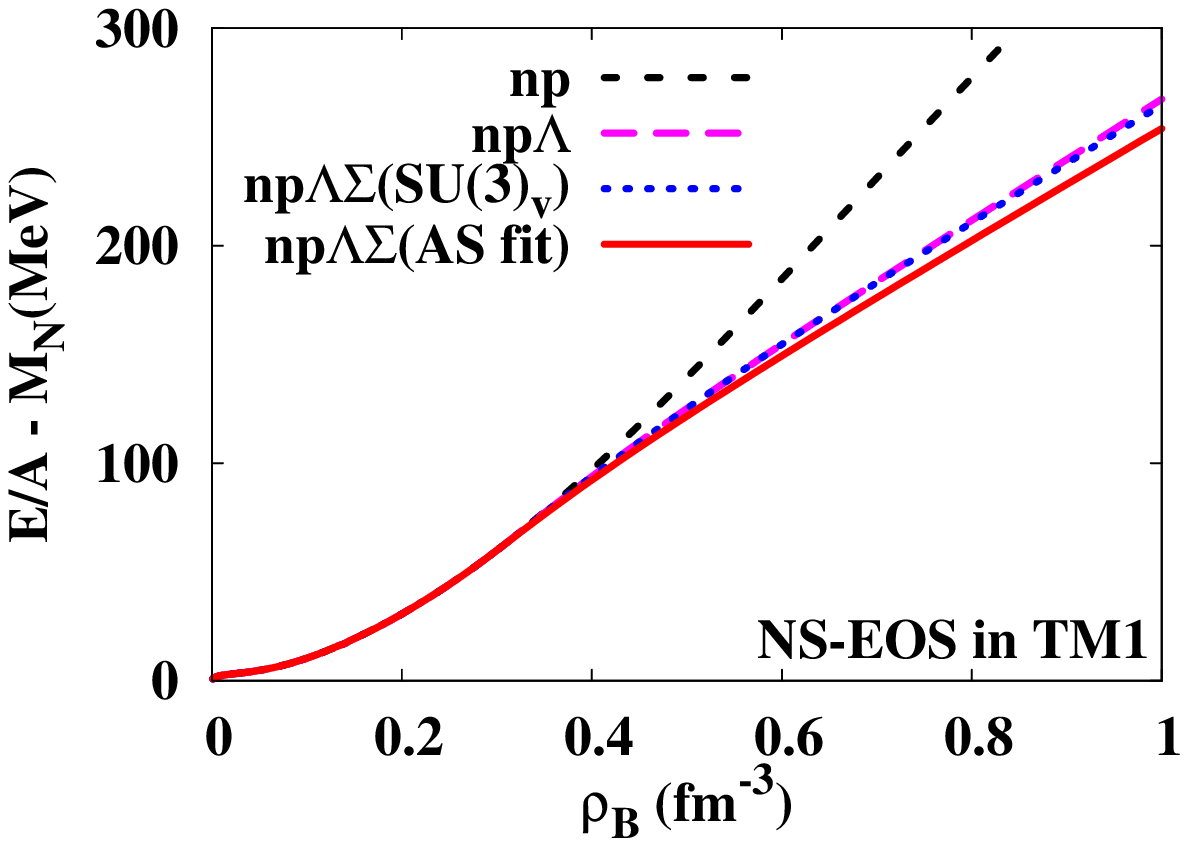}
	\end{center}
	\caption{
		Calculated NS-EOS based on the RMF models with $\Lambda$ and
		$\Sigma$ hyperons.
		Upper panel shows NS-EOS in the NL1, NL-SH, TM1 and SCL2 RMF
		models those coupling constants to hyperons have been
		determined so as to reproduce $\Lambda$ and $\Sigma$
		hypernuclear data in this article (AS fit values).
		Broken lines correspond to NS-EOS with $n$ and $p$
		and solid lines represent NS-EOS including $\Lambda$ and
		$\Sigma$ hyperon additionally. 
		Leptons ($e$ and $\mu$) are also considered.
		In lower panel, we compare NS-EOSs with \SUv\ and AS fit values
		in the TM1 RMF model.
		}
	\label{fig:NSEOS}
\end{figure}

In Fig.~\ref{fig:NSEOS}, we compare energy per baryon, $E/A - m_n$,
in NS matter in NL1, NL-SH, TM1 and SCL2 parameter sets
as a function of baryon density, $\rhoB$.
Compared to TM1 and SCL2, NL1 and NL-SH give us stiffer NS-EOSs.
By including the hyperon effects, all of NS-EOSs are significantly softened.
In the bottom panel of Fig.~\ref{fig:NSEOS},
we compare the NS-EOSs in several cases in TM1,
nucleon ($np$) matter, $np\Lambda$ matter, 
and $np\Lambda\Sigma$ matter with \SUv\ and AS fit values of $g_{\rho\Sigma}$.
We find that the emergence of $\Sigma^-$ hyperon softens
NS-EOSs further but slightly if we adopt AS fit values of $g_{\rho\Sigma}$.
By comparison, the NS-EOSs with \SUv\ values
are almost the same as NS-EOSs composed of $np\Lambda$.
It is a general trend that
NS-EOS suggested by fitting the $\Sigma^-$ atomic shifts
becomes slightly softer than the \SUv-constrained EOS.

\begin{figure}[tb]
	\begin{center}
		\includegraphics[width=0.48\textwidth]{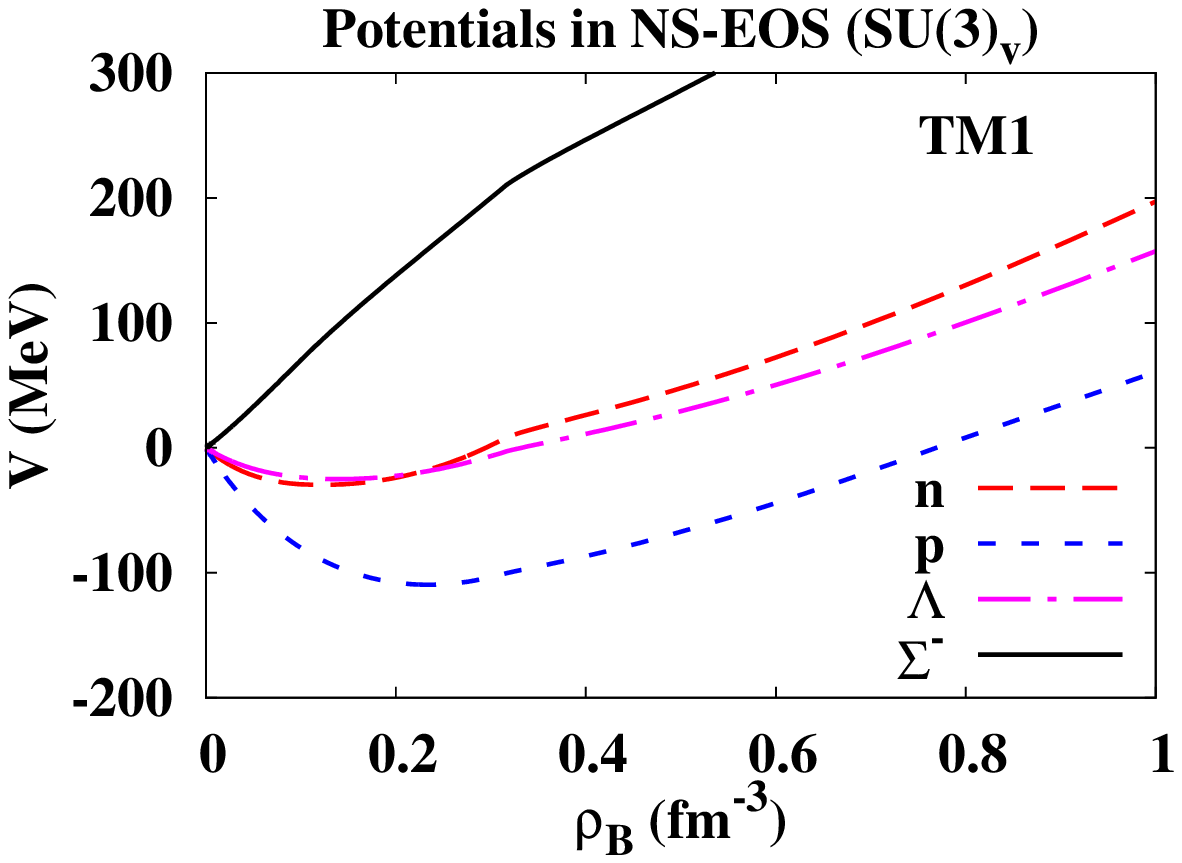}
		\includegraphics[width=0.48\textwidth]{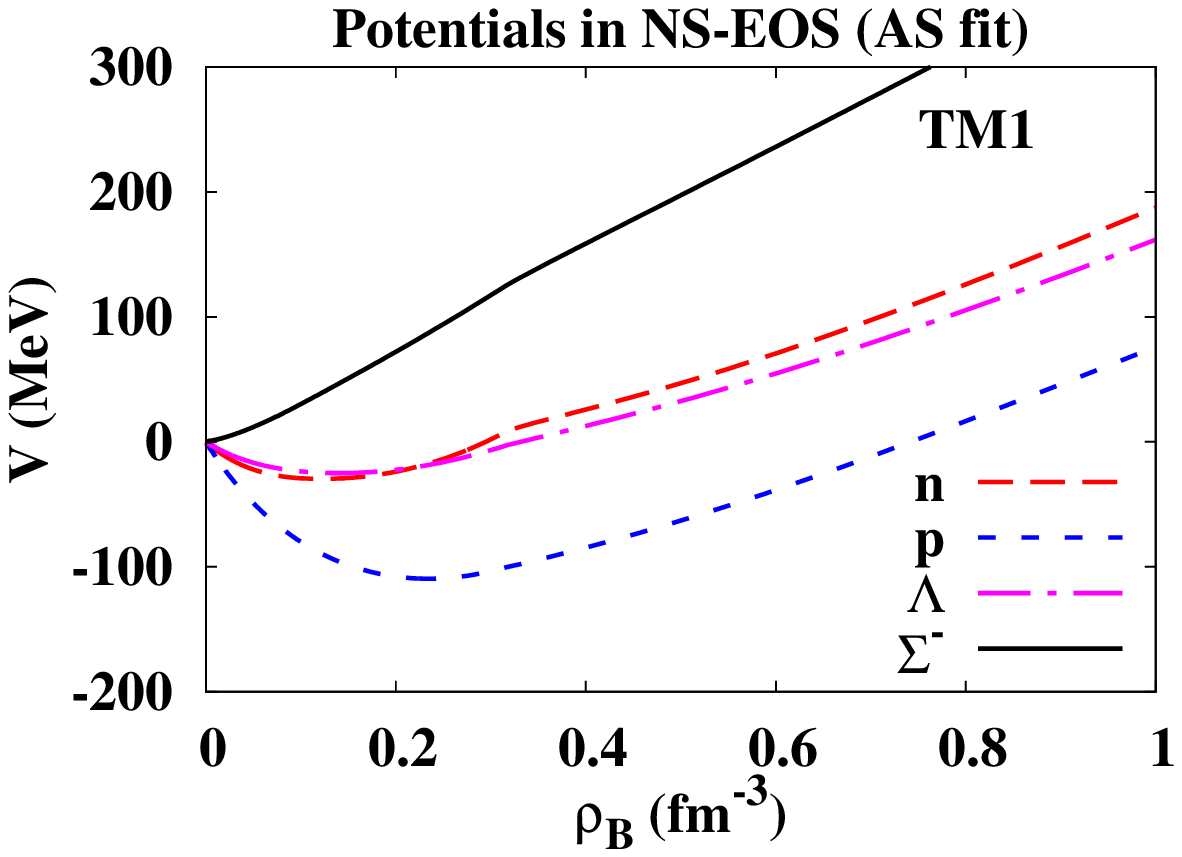}
	\end{center}
	\caption{Calculated potentials of baryons in NS matter on TM1 RMF
		model.
		Upper panel shows the results with \SUv\ $g_{\rho\Sigma}$
		values, Eq.~(\ref{Eq:SUV3}).
		Lower panel displays the ones with $g_{\rho\Sigma}$ determined
		by reproducing $\Sigma^-$ atomic shifts (AS fit values).
		}
	\label{fig:potNSoC_P}
\end{figure}

To confirm these results, we examine baryon potentials in NS matter,
which are shown in Fig.~\ref{fig:potNSoC_P}.
Compared to \SUv\ cases,
it is clear that the $\Sigma^-$ potentials with AS fit values
become less repulsive.
This less repulsive $\Sigma^-$ potential
may allow $\Sigma^-$ hyperon to appear in NS matter with AS fit values
and soften NS-EOS.
%
 
\begin{figure}[tb]
	\begin{center}
	\includegraphics[width=0.5\textwidth]{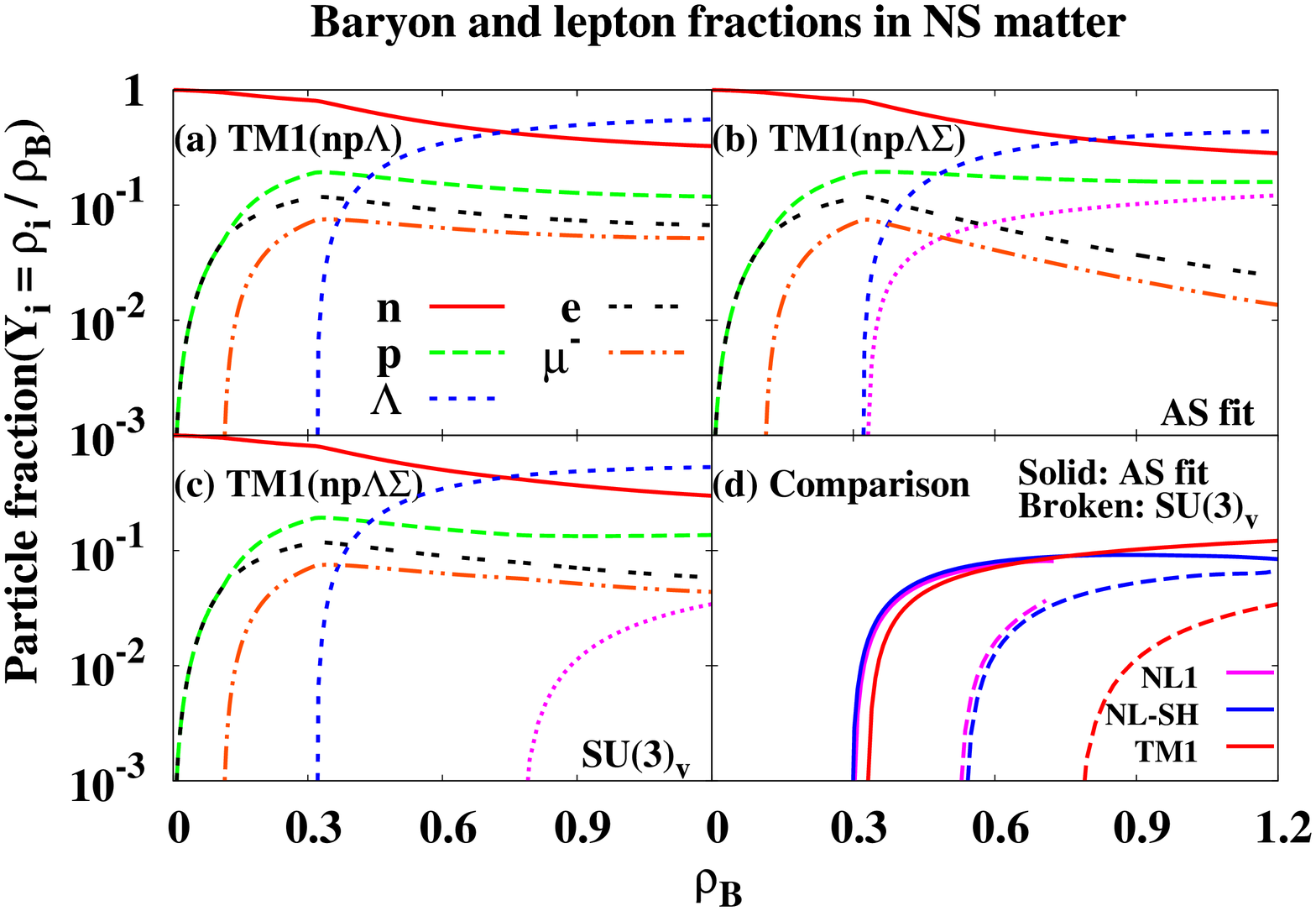}
	\end{center}
	\caption{
	        Baryon and lepton fractions, $Y_B$ and $Y_L$ calculated based
		on applied RMF model in which hyperon degrees of freedom are
		introduced.
		(a) $np\Lambda$ NS matter, (b) $np\Lambda\Sigma$ NS matter 
		($g_{\rho\Sigma}$: fixed by reproducing the atomic shifts of
		$\Sigma^-$), and (c) $np\Lambda\Sigma$ NS matter
		($g_{\rho\Sigma}$: constrained by \SUv\
		symmetric relation) based on TM1 RMF model, respectively;
		(d) comparison among the results of NL1, NL-SH, and TM1 RMF
		models.
		}
	\label{fig:Y_B}
\end{figure}

We show baryon and lepton fractions, $Y_{B,L}$, in NS matter with TM1 
for several choices of hyperon effects in Fig.~\ref{fig:Y_B}.
In Figs.~\ref{fig:Y_B} (b) and ~\ref{fig:Y_B} (c),
we show calculated baryon and lepton fractions
with AS fit and \SUv\ values of $g_{\rho\Sigma}$, respectively.
From these results, if we apply AS fit $g_{\rho\Sigma}$,
$\Sigma^-$ tend to appear in NS matter from lower $\rho_B$ compared to
that with \SUv\ value.
This trend is already suggested in our previous work using SCL3 RMF.
Thus, we confirm that it is common to all employed RMF models here
and SCL3 RMF model that $\Lambda$ and $\Sigma^-$ appear
almost simultaneously at around 2-3 $\rho_0$
if we apply the parameter sets reproducing $\Sigma^-$ atomic shifts,
as shown in Fig.~\ref{fig:Y_B} (d).

\begin{figure}[tb]
	\begin{center}
		\includegraphics[width=0.5\textwidth]{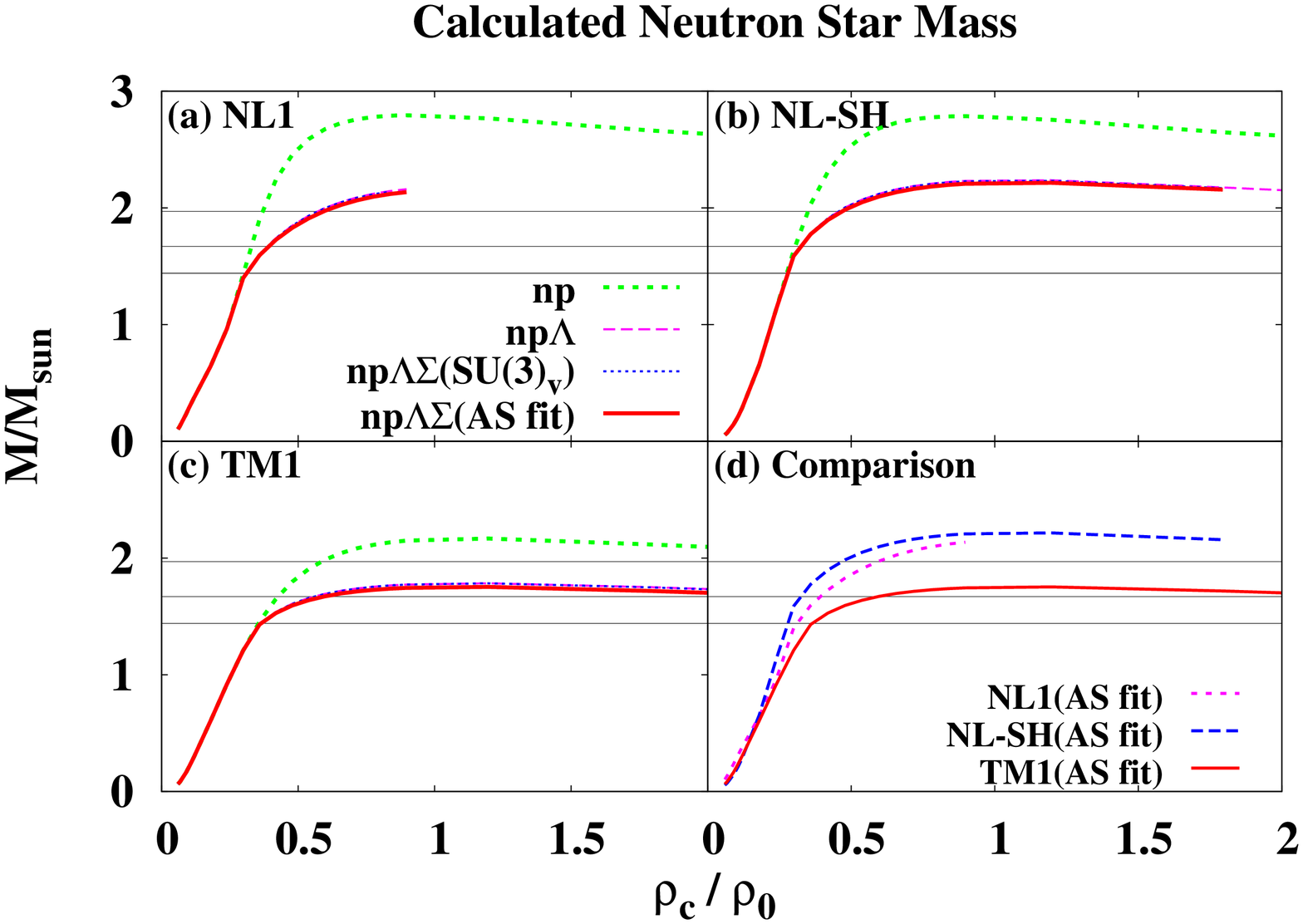}
	\end{center}
	\caption{
		Calculated NS mass by solving TOV equation based on the
		RMF model parameters whose meson-hyperon couplings
		are determined in these references,
		(a)NL1, (b)NL-SH, and (c) TM1 RMF models, respectively.
		In panel (d), the comparison among the results of NL1, NL-SH,
		and TM1 RMF models is presented with green dash line,
		blue dotted line, and red solid line, respectively.
		}
	\label{fig:NLRMFTOV}
\end{figure}
In Fig.~\ref{fig:NLRMFTOV}(a)-(c), we show calculated NS mass 
in NL1, NL-SH and TM1 as a function of central baryon density $\rho_c$,
respectively.
Maximum masses of hyperonic stars are reduced
by including $\Sigma^-$ in all models.
At the same time, NS-EOSs have already been softened strongly
by the emergence of $\Lambda$ hyperon,
and the softening effect of $\Sigma^-$ on NS maximum mass is not very strong.
In NL1 and NL-SH,
calculated maximum masses of NS exceed 2$M_\odot$ 
even if $\Lambda$ and $\Sigma$ hyperons are included,
and the recently observed heavy NS~\cite{NSmass2} can be supported.
By comparison,
the calculated NS maximum mass in TM1 with hyperons
does not reach 2$M_\odot$,
and the two-solar-mass NS puzzle remains.
%

It would be premature to conclude that 
the two-solar-mass NS puzzle can be solved 
in NL1 and NL-SH RMF models with AS fit values of $g_{\rho\Sigma}$.
It seems that the high density region in NS core
may be out of the range of applicability 
in the present treatment of NL1 and NL-SH parameter sets with hyperons;
the effective mass of nucleon is reduced too much and it becomes negative
at around 4$\rho_B$ and 6.5$\rho_0$, respectively.
The mechanism of the negative nucleon effective mass can be understood
as follows.
Both nucleons and hyperons act to increase $\sigma$
as long as their effective mass is positive.
At the density where nucleon effective mass vanishes,
hyperons are still massive
due to the smaller couplings with $\sigma$ and larger masses in vacuum.
These huge mass reductions may correspond to the phase transition
from a baryonic matter to a quark matter since they indicate the complete
restoration of chiral symmetry.

By comparison, TM1 parameter set is free from the negative nucleon mass
problem in the density region considered here,
but its maximum mass lies below the observed $1.97M_\odot$;
maximum mass of NS in TM1 with hyperons is calculated to be 1.75$M_\odot$.
In TM1, the $\omega^4$ self-interaction is introduced so as
to simulate the scalar and vector potentials
in Dirac-Br{\"u}ckner-Hartree-Fock (DBHF) calculation~\cite{DBHF}.
We have also adopted this $\omega^4$ self-interaction in SCL2 and SCL3 models.
The $\omega^4$ term suppresses $\omega$ field at high density,
and softens the EOS.
Then a model with a larger coefficient $c_{\omega 4}$
predict a smaller NS maximum mass.
Since NL RMF models do not include $\omega^4$ terms,
vector repulsive potentials linearly increase at higher $\rhoB$
as we presented in Fig.~5 of Ref.~\cite{SCL3}.

Different predictions in the RMF models discussed here
implies the importance of three-body interactions~\cite{Nishizaki:2002ih}.
Higher order meson interaction terms such as $\sigma^{3,4}$ and $\omega^4$
may be related to the three-body interactions.
These interactions are expected to lead not only re-stiffening effect 
in EOS but also the suppression to the appearance of hyperons.
We introduced explicit three-body couplings to RMF model and examined
their effects to high density NS-EOS with preliminary parameter sets
which are determined in the same way as we have reported in this work
~\cite{TO13}.
More detail analysis is strongly needed to give a conclusion of 
the validity of three-body couplings.

\section{Summary and discussion}

In this article, we have investigated neutron star matter equation of state
(NS-EOS) based on the RMF models which can reproduce
the bulk properties of nuclear systems consisting of nucleons
and hypernulcear systems 
consisting of $\Lambda$ and $\Sigma$ hyperons in addition to nucleons.
We have emphasized the importance of $\Sigma$ coupling
with isovector-vector ($\rho$) meson, $g_{\rho\Sigma}$.

The coupling constants of $\Lambda$ and $\Sigma$ hyperons in RMF models
have been well-constrained by explaining 
separation energies of $\Lambda$ ($S_\Lambda$)
in single $\Lambda$ hypernuclei,
$\Lambda\Lambda$ bond energy ($\Delta B_{\Lambda\Lambda}$)
in the double $\Lambda$ hypernucleus
$^{6}_{\Lambda\Lambda}\mathrm{He}$,
and the $\Sigma^-$ atomic shifts,
under the assumption that the isoscalar vector couplings are fixed
by the \SUv\ symmetric coupling constant relations.
We have found that we need to reduce $g_{\rho\Sigma}$ from the \SUv\ value
to explain the experimental $\Sigma^-$ atomic shifts.
%
Then the symmetry energy of $\Sigma$ in the atomic shift fit case
is smaller than that in the \SUv\ case.
%
%

We have examined the hyperon effects on the NS-EOS
by using the RMF including hyperons
with AS fit and \SUv\ values of $g_{\rho\Sigma}$.
We have confirmed that $\Sigma^-$ 
would appear in NS matter with the AS fit $g_{\rho\Sigma}$ value,
since the isovector part of $\Sigma^-$ potential in NS matter is smaller
in the AS fit case than in the \SUv\ case.
This trend is common to all of employed RMF parameter sets, 
NL1, NL-SH, TM1 and SCL2 with AS fit values.
$\Sigma^-$ hyperons can appear in NS matter at $\rhoB \sim 2\rho_0$, 
which is close to the density where $\Lambda$ hyperons appear.
Thus, it is valuable to revisit the appearance of $\Sigma^-$ in NS matter,
when we want to understand NS based on experimental hypernuclear
data including $\Sigma^-$ atomic shifts.

NS-EOS is softened by $\Lambda$ and $\Sigma$ hyperons.
The NS maximum mass is reduced
by $(0.5-0.6) M_\odot$ when we include $\Lambda$, 
and it is further reduced by including $\Sigma$ slightly.
When we include hyperons in NS-EOS,
TM1 RMF model cannot support $2 M_\odot$ NS~\cite{NSmass2, NSmass3},
while NL1 and NL-SH could support 2$M_\odot$ NS mass.
We need more studies to solve the heavy neutron star puzzle conclusively.
In NL1 and NL-SH models,
nucleon effective mass becomes negative at medium baryon density,
and we cannot explain the density dependence of the vector potential 
obtained in the Dirac-Br{\"u}ckner-Hartree-Fock (DBHF) calculation.
In TM1, SCL2, and SCL3 models,
$\omega^4$ self-interaction is included to simulate the DBHF results,
but these models cannot support the 2$M_\odot$ NS.
 
It was also suggested that $\Xi^-$-nucleus potential is attractive
from the analysis about $\Xi^-$ production spectrum~\cite{XiData}.
If $\Xi$ hyperons emerge in dense matter, NS-EOS will be softened a little more
as in the $\Sigma^-$ case.
We guess, however, that maximum mass will not be affected by inclusion of
$\Xi^-$ so much since, as we have shown, NS-EOS has already been softened
by including $\Lambda$ hyperon and $\Sigma^-$ reduce the electron chemical
potential.

From these results, we conclude that re-stiffening mechanisms are required
to understand the massive NS properties.
As one of candidate to solve this problem,
three-body repulsive interactions are suggested to be considered.
In Ref.~\cite{Nishizaki:2002ih},
universal three-baryon repulsive interactions were examined.
These interactions are expected to lead not only re-stiffening effect but also
the suppression to the appearance of $\Lambda$, $\Sigma$ and $\Xi$ hyperons.
Thus, further investigation is needed 
and preliminary results with explicit three-body couplings in RMF model
are reported in Ref.~\cite{TO13}.
The detailed analysis
is in progress and will be reported elsewhere.

\begin{acknowledgments}
This work was supported in part by
the Grants-in-Aid for Scientific Research from JSPS
(Nos.
	(B)23340054 (A.~Nakamura(incl. A.~Ohnishi))	,
	(B)23340067 (T.~Kunihiro(incl. A.~Ohnishi))     ,
	(C)23340271 (A.~Ohnishi, K.~Morita, T.~Kunihiro),
    and (C)25400278, (T.~Harada)
),
by the Grants-in-Aid for Scientific Research on Innovative Areas from MEXT
(No. 2404: 24105001, 24105008), 
by the Yukawa International Program for Quark-hadron Sciences,
and by the Grant-in-Aid for the global COE program ``The Next Generation
of Physics, Spun from Universality and Emergence'' from MEXT.
\end{acknowledgments}

\end{document}